\documentclass[aps,showpacs,twocolumn,prl]{revtex4-1}
\usepackage{graphicx,epsfig}
\usepackage{dcolumn}
\usepackage{bm}
\usepackage{psfrag}
\usepackage{float}
\usepackage{plain}
\usepackage{tabularx}
\usepackage[latin1]{inputenc}
\usepackage{amsmath}
\usepackage{verbatim} 
\begin{document}

\title{Reply to the comment on:\\
``Thermostatistics of Overdamped Motion of Interacting Particles''\\
 by Y. Levin and R. Pakter}

\author{J.~S. Andrade Jr.$^{1,3}$, G.~F.~T. da Silva$^{1}$, A.~A.
  Moreira$^{1}$, F.~D. Nobre$^{2,3}$, E.~M.~F. Curado$^{2,3}$}

\affiliation{$^{1}$Departamento de F\'{\i}sica, Universidade Federal
  do Cear\'a, 60451-970 Fortaleza, Cear\'a, Brazil\\ $^{2}$Centro
  Brasileiro de Pesquisas F\'{\i}sicas, Rua Xavier Sigaud 150,
  22290-180, Rio de Janeiro-RJ, Brazil\\$^{3}$National Institute of
  Science and Technology for Complex Systems, Rua Xavier Sigaud 150,
  22290-180, Rio de Janeiro-RJ, Brazil}

\maketitle

In their comment, Levin and Pakter present an analytical solution for
the mechanical equilibrium of a system of particles interacting
through a potential given by the modified Bessel function, and
confined by a restoring force. This derivation yields a quadratic
dependence for the local density of particles, which is consistent with
Eq.~(14) of our work. Based on this result, Levin and Pakter question
our interpretation that a system of overdamped particles at T=0 is a
physical realization of Tsallis thermostatics with entropic index
$\nu=2$. They claim that the same density profile can be obtained
using only Newton's laws.

\vskip 0.25cm

In what follows we provide a reply to the remarks of Levin and Pakter.
We first show that their attack to our results and analysis is
conceptually unfounded and rather misleading. Inexplicably, they
simply choose to categorically dismiss our elaborated and solid
conceptual approach and results, without employing any fundamental
concepts or tools from Statistical Physics. We then demonstrate that
the results of Levin and Pakter do not present any evidence against,
but rather corroborates, our conclusions. In fact, the results shown
in their comment correspond to a confining potential that is $1000$
times stronger than the typical valued utilized in our study,
therefore explaining the discrepancy between their results and ours.
Furthermore, in this regime where higher vortex densities are
involved, vortex cores might get so close to each other that can no
longer be treated as point-like defects. As a consequence,
Ginzburg-Landau equations should be employed instead, meaning that the
physical conditions implied by the results of Levin and Pakter should
be considered with caution in the context of the Physics of
interacting superconducting vortexes.

\vskip 0.25cm

{\bf 1) Levin and Pakter claim that our result for the density
  distribution of particles ``{\it has nothing to do with the Tsallis
    statistics}''.}

\vskip 0.1cm

In our letter, we show that the the non-linear Fokker-Planck equation
in the form
\begin{equation}
\label{eq:tsallisnlfpe}
\frac{\partial P}{\partial t}= -\frac{\partial[A(x)P]}{\partial x}
+Dq \frac{\partial}{\partial x}
\left[ P^{q-1} \frac{\partial P}{\partial x} \right]~,
\end{equation}
results in a dynamics that drives the system towards a state that
maximizes Tsallis entropy with entropic index $\nu$. See also~\cite{nlfpe}
where this is discussed in more detail. Note that Levin and Pakter do not
contest this statement.

We also show that, for a class of systems of overdamped particles
interacting through repulsive short range potentials, the
particle-particle interaction induces a flux of particles proportional
to gradient of concentration. This has been previously discussed in
few other works~\cite{cm}. We also show in our letter that this kind of
dynamics leads to a non-linear diffusion equation that is equivalent
to the non-linear Fokker-Planck equation~(\ref{eq:tsallisnlfpe}). Again,
none of these statements were contested by Levin and Pakter.

In view of the aforementioned facts we are rather surprised that Levin
and Pakter, without contesting any of the results demonstrated in our
work, still question our conclusion that ``{\it a system of overdamped
  particles at T=0 is a physical realization of Tsallis thermostatics
  with entropic index $\nu=2$.}'' This is specially puzzling when one
notes that the solution proposed by Levin and Pakter for this system
also corresponds to a solution of our non-linear equation, therefore
confirming our findings.

\vskip 0.25cm

{\bf 2) Levin and Pakter state that our result for the density profile
  of particles ``{\it has everything to do with Newton's Second
    Laws}''.}

\vskip 0.1cm

Of course, we were by no means expecting our results to be in
contradiction with Newton's Laws. As a matter of fact, in our study,
we used molecular dynamics to integrate equations of motion and obtain
numerically density profiles, validating the predictions of our
non-linear Fokker-Planck equation. In the same way, neither the
Maxwell-Boltzmann nor the Tsallis thermostatistics are contrary to
classical mechanics theory, as one could infer from the final and
rather misleading sentence of the comment.

\vskip 0.25cm

{\bf 3) In their comment, Levin and Pakter also state that ``{\it
    prior to discarding the standard statistical mechanics one should
    see what it has to say on this matter.}''}

\vskip 0.1cm

Again, our approach and results neither disregard nor contradict
``standard'' statistical mechanics. We would surely welcome any other
treatment based on the solid principles of Statistical Physics. This
is definitely not the case of this comment. Instead, however, they
categorically dismiss our elaborated approach and results without
employing any concept or tool of Statistical Physics.

\vskip 0.25cm

{\bf 4) Levin and Pakter write in their comment that ``if a classical
  system is placed in contact with a temperature reservoir at $T=0$ it
  will loose all its kinetic energy and collapse to the ground state''
  where the ``net force on each particle vanishes.'' They also state
  in their comment that ``this is precisely what happens for the
  system studied by Andrade et al.''}

\vskip 0.1cm

Of course, the overdamped motion we investigate should evolve towards
mechanical equilibrium, as we explicitly state in our paper.  One
should note, however, that our approach goes beyond the prediction of
a stationary state, since the non-linear Fokker-Planck equation
introduced by us also describes the dynamics of the system, with all
its transient features, till the stationary regime is eventually
reached. The analytical solution provided in the comment, however,
does not contemplate the dynamics outside the stationary state. The
fact that our system evolves towards mechanical equilibrium is in no
way contradictory to our results or conclusions.

It is important to point out that we also studied the case in which 
thermal noise disrupts equilibrium, resulting in a density function
that is consistent with a mixed entropy combining Tsallis with entropic
index $\nu=2$, and Boltzmann-Gibbs formulations, namely, Eq.~(17) of
our paper. By varying the ratio between the particle-particle
interaction and thermal force, we can continuously interpolate the
density profile from a Gaussian, at higher temperatures, to a
parabolic shape at $T=0$. In short, what we proposed in our paper and
the authors of the comment appeared to miss completely, is that
overdamped particles interacting through a wide family of potentials
follow a dynamics that drives the system to the maximum value of
the entropic function Eq.~(17) of our paper. The usefulness of our
approach is that the density profile can be easily obtained with the
many-body interactions being carried into the entropic term with
$\nu=2$.

\begin{figure}[t]
\includegraphics*[width=8cm]{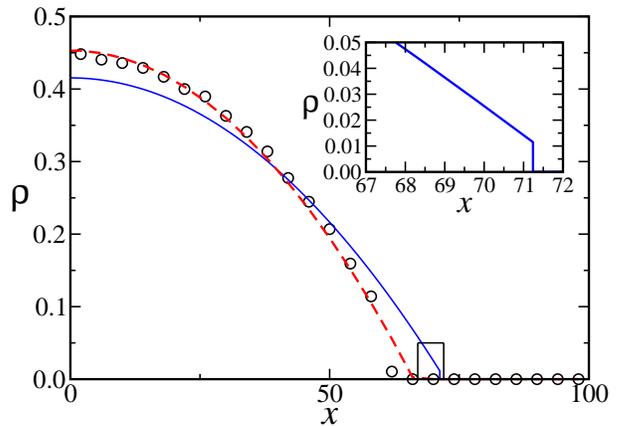}
\caption{Numerical results for the density profile obtained from
  numerical simulations. We used the same conditions as Fig.~1 of our
  letter, namely, $N=800$,$\alpha=10^{-3}f_0$, $L_y=20\lambda$. The
  dashed red line is the solution for this profile given by Eq.~(14)
  of our letter, while the straight blue line correspond to the
  solution according to Eq.~(3) from Levin and Pakter's comment.
  Clearly their solution is not compatible with the concavity of the
  density profile obtained from our molecular dynamics simulations.
  The inset shows that the discontinuity in their solution is barely
  noticeable for these physical conditions.}
\label{f.1}
\end{figure}

\vskip 0.25cm  

{\bf 5) In their comment Levin and Pakter claim that they obtain an
  exact solution for the particle distribution at $T=0$.}

\vskip 0.1cm

This claim is misleading to say the least. In fact, in the same way as
we did, their solution is a coarse-grain approximation, where the
discrete nature of the interacting particles is replaced by a
continuum description for the particle density. However, it is
relevant to note that they arrive at a solution that is similar, but
not identical to ours. Levin and Pakter also show results from
numerical simulations that seem to agree with their solution. As we
show here, however, they studied the system at a regime that is
significantly different from the one we investigated in our work, and
that is clearly the reason for the discrepancies.

The equation of motion for our system of overdamped particles is given by
\begin{equation}
\mu \vec{v}_{i}=\sum_{j \neq i}f_0K_1(|\vec{r}_{ij}|/\lambda)\hat{r}_{ij}-
\sum_i{\alpha x_i}
\label{eq:mov}
\end{equation}
where $K_1$ is the modified Bessel function. The factors $f_0$ and
$\alpha$ control the intensity of the particle-particle interaction
and the confining force, respectively. In their results Levin and
Pakter used $\alpha=f_0$ while we used $\alpha=10^{-3}f_0$. It is not
surprising that, by using a confining potential $1000$ times stronger,
one should obtain results that are distinct from ours.

\vskip 0.25cm

{\bf 6) Levin and Pakter arrive at a result that is similar to
  Eq.~(14) of our paper, namely, Eq.~(3) of their comment. They claim,
  however, that their solution does not need any adjustable parameter,
  while we had to set the value of the parameter that controls the
  concavity of the distribution to $a=2.41f_o\lambda$, in order to
  adequately fit our model to our results from molecular dynamics
  simulations.}

\vskip 0.1cm

Apparently the authors of the comment did not realize that their
proposal of $a=\pi f_0\lambda^3$ as a ``non-adjustable parameter''
corresponds exactly to the same theoretical prediction presented in
our paper. However, this prediction is valid only when the density of
particles $\rho$ varies slowly within the interaction range of the
potential. Since the system is composed of discrete particles, the
value of the parameter $a$ should approach $\pi f_0\lambda^3$ as the
density grows. Since Levin and Pakter use a substantially stronger
confining potential, the obtained densities can reach $5$ particles
per $\lambda^2$, which is about ten times higher than ours. At this
regime, the value $a=\pi f_0\lambda^3$ represents indeed a good fit to
the numerical data. At moderates densities, however, the numerical
results do not follow this prediction. In Fig.~\ref{f.1}, we show our
numerical results and parabolic predictions with $a=2.41 f_0\lambda^3$
and $a=\pi f_0\lambda^3$. It is clear that the value of $a$ used by
Levin and Pakter do not agree with numerical results from molecular
dynamics in this regime.

In any case, it is important to emphasize that the interacting
potential in the form of the modified Bessel function is originally
motivated by applications in the theory of superconducting vortexes.
It is a known fact~\cite{sc} for this physical system that such a
potential form represents an approximated model that is only valid in
the regime of moderate magnetic fields (i.e., moderate particle
densities, $\rho \ll \kappa^{2}/2 \pi \lambda^{2}$, where $\kappa$ is
the Ginzburg-Landau parameter). In the case of higher densities, the
vortex cores might get so close to each other that can no longer be
treated as point-like defects. As a consequence, Ginzburg-Landau
equations should be employed instead, meaning that the physical
conditions implied by the results of Levin and Pakter should be
considered with caution in the context of the Physics of interacting
superconducting vortexes.

\vskip 0.25cm

{\bf 7) Levin and Pakter predict the existence of discontinuities at
  the edges of the density function $\rho$.}

\vskip 0.1cm

We would like to point out that, although the numerical results of
Levin and Pakter follow their theoretical prediction in the bulk of
the density profile, by just looking at their histogram it is not
clear whether the discontinuity is really present of not. The inset of
Fig.~\ref{f.1} shows that under the conditions of our simulations,
this discontinuity, if present, would be negligible. Therefore, we
conclude that, in the conditions we model our system, the
discontinuity predicted by Levin and Pakter does not represent a
relevant effect. In any case, the analytical result of Levin and
Pakter can be viewed as a solution of Eq.~(13) of our letter, with
$a=\pi f_0\lambda^3$, but subjected to a different boundary condition.
Therefore, our understanding is that their results are also consistent
with our theoretical model.

\vskip 0.25cm

In summary, based on the preceding comments and remarks, it is our
opinion that the comment by Y. Levin and R. Pakter is neither
conceptually sound nor relevant.

\end{document}